\RequirePackage{lineno}
\documentclass[twocolumn,superscriptaddress,preprintnumbers,amsmath,amssymb,prc,nofootinbib]{revtex4}  
\usepackage{graphicx}
\usepackage{dcolumn}
\usepackage{bm}
\usepackage{float}
\usepackage{color}
\usepackage{CJK}
\usepackage{tikz}
\usetikzlibrary{shapes,arrows}
\usepackage[colorlinks,linkcolor=blue,urlcolor=blue,citecolor=blue]{hyperref}
\usepackage{isotope}
\usepackage[normalem]{ulem}
\renewcommand{\sout}{\bgroup \color{red} \ULdepth=-0.5ex \ULset}

\usepackage{amsmath}
\usepackage{cases}
\usepackage{amssymb}
\usepackage{eqnalign}

\begin{document}
\begin{CJK*}{UTF8}{gbsn}

\title{Probing fluctuations and correlations of strangeness by net-kaon cumulants in Au+Au collisions at $\sqrt{s_{NN}} = 7.7$ GeV}

\author{Qian Chen}
\affiliation{Shanghai Institute of Applied Physics, Chinese Academy of Sciences, Shanghai 201800, China}
\affiliation{University of Chinese Academy of Sciences, Beijing 100049, China}
\affiliation{Key Laboratory of Nuclear Physics and Ion-beam Application (MOE), Institute of Modern Physics, Fudan University, Shanghai 200433, China}
\affiliation{Shanghai Research Center for Theoretical Nuclear Physics, NSFC and Fudan University, Shanghai $200438$, China}
\author{Han-Sheng Wang}
\affiliation{Key Laboratory of Nuclear Physics and Ion-beam Application (MOE), Institute of Modern Physics, Fudan University, Shanghai 200433, China}
\affiliation{Shanghai Research Center for Theoretical Nuclear Physics, NSFC and Fudan University, Shanghai $200438$, China}
\author{Guo-Liang Ma}
\email[]{glma@fudan.edu.cn}
\affiliation{Key Laboratory of Nuclear Physics and Ion-beam Application (MOE), Institute of Modern Physics, Fudan University, Shanghai 200433, China}
\affiliation{Shanghai Research Center for Theoretical Nuclear Physics, NSFC and Fudan University, Shanghai $200438$, China}
\affiliation{Shanghai Institute of Applied Physics, Chinese Academy of Sciences, Shanghai 201800, China}
	
	
\begin{abstract}
We calculate the cumulants and correlation functions of net-kaon multiplicity distributions in Au+Au collisions at $\sqrt{s_{NN}} = 7.7$ GeV using a multiphase transport model (AMPT) with both a new coalescence mechanism and all charge conservation laws. The AMPT model can qualitatively describe the centrality dependences of the net-kaon cumulants and cumulant ratios measured by the STAR experiment. By focusing on the stage evolution of the cumulants, cumulant ratios, and correlation functions, we reveal several key effects on the fluctuations and correlations of strangeness during the dynamical evolution of relativistic heavy-ion collisions, including strangeness production and diffusion, hadronization, hadronic rescatterings, and weak decays. Without considering the quantum chromodynamics critical fluctuations in the dynamic model, we demonstrate that the net-kaon fluctuations can largely represent the net-strangeness fluctuations. Our results provide a baseline for understanding the net-kaon and net-strangeness fluctuations, which help to search for the possible critical behaviors at the critical end point in relativistic heavy-ion collisions.

\end{abstract}
	
\maketitle
	
\section{Introduction}
\label{framework}

\label{introduction}
The study of quantum chromodynamics (QCD) phase diagram has been a topic of great interest for many years~\cite{Cabibbo:1975ig,Fukushima:2010bq,Bzdak:2019pkr}, which aims to show the conditions at which thermodynamically distinct phases of strongly interacting matter occur or coexist at equilibrium. It has been shown that the phase transition at the vanishing baryon chemical potential ($\mu _{B}$) is a smooth cross over~\cite{Aoki:2006we} with a pseudocritical temperature of $T_{C} \simeq 160$ MeV~\cite{Aoki:2009sc,Bazavov:2011nk} from the lattice QCD with the first principle~\cite{Ding:2015ona,HotQCD:2014kol,Bellwied:2013cta,Borsanyi:2014ewa}. On the other hand, some effective QCD models predicted the phase transition to be first order at large $\mu _{B}$~\cite{Ejiri:2008xt,Alford:1997zt}. The end point of the first-order phase boundary toward the crossover region is called the QCD critical end point (CEP)~\cite{Skokov:2010uh,Aoki:2006we,Karsch:2011gg,Stephanov:2004wx}. Many efforts have been made by both the experimental and theoretical communities to find the location of the CEP~\cite{Jeon:2000wg,Koch:2008ia,Bzdak:2018uhv,Chatterjee:2015oka,Mitter:2014wpa,Herbst:2013ufa,Luo:2017faz,Fu:2018qsk,Fu:2019hdw,Fukushima:2003fw,Skokov:2010uh,Pisarski:2016ixt,Li:2017ple,Chen:2018vty,Shao:2016pvg,Liu:2013imm}. The fluctuations of conserved charges have been proposed as a promising observable to probe the CEP~\cite{Stephanov:2008qz,Stephanov:2011pb,Gupta:2011wh}, because they are very sensitive to the diverged correlation length~\cite{Stephanov:2008qz,Athanasiou:2010kw,Stephanov:2011pb,Cheng:2008zh,Gavai:2010zn,Chen:2021kjd,Xu:2015hrj} and directly related to the susceptibilities in the lattice QCD calculations~\cite{Gupta:2011wh,Ding:2015ona,Bazavov:2012vg}.

The phase structure of strongly interacting QCD matter can be achieved by performing heavy-ion collision experiments at different energies~\cite{STAR:2005gfr,STAR:2010mib,BRAHMS:2004adc,Friman:2011pf}. The range of chemical freeze out points extracted from the heavy-ion collision experiments covers from the almost vanishing baryon chemical potential at the CERN Large Hadron Collider (LHC) to about $\mu_{B}\simeq 750$ MeV~\cite{STAR:2022vlo} at the BNL Relativistic Heavy Ion Collider (RHIC). The RHIC has been performing a beam energy scan (BES) program since 2010~\cite{STAR:2010vob,STAR:2017sal,STAR:2010mib,STAR:2014egu,STAR:2013gus}. The RHIC-STAR BES-I experiments measured the cumulants of net-proton (proxy of net-baryon), net-charge, and net-kaon (proxy of net-strangeness) multiplicity distributions in Au+Au collisions at the different center of mass of energies from 7.7 up to 200 GeV~\cite{STAR:2014egu,STAR:2013gus,Pandav:2020uzx,Thader:2016gpa}. Recently, the RHIC experiments have predicted that the critical phenomenon is likely to occur at $3<\sqrt{s_{NN}}<20$ GeV based on the measurements of net-proton high order cumulants~\cite{STAR:2021fge}. 

Since the incident nuclei in heavy ions do not carry strangeness, strange hadron production in the final state can provide a unique way to understand the characteristics of the created system. For example, strange quarks were predicted to be produced abundantly in the quark-gluon plasma (QGP), which leads to the enhancement of strange hadrons, one of the well-known signatures of QGP formation~\cite{Rafelski:1982pu}. Another famous experimental signal is the peak of the energy dependence of $K^+/\pi^+$ ratio observed in the central Pb+Pb collision at $30A$ GeV, which is thought as the possible onset of the QCD deconfinement~\cite{NA49:2007stj}. The enhancement of strange baryons has been observed in high-energy $A+A$ collisions relative to $p+p$ collisions because of the enhanced yield of multi-strange particles produced in the presence of QGP~\cite{Antinori:2002qr,STAR:2007cqw}. The yields and fluctuations of different kinds of hadrons can reveal the information about chemical freeze-out in heavy-ion collisions. The analyses of the experimental data showed that light hadrons seem to prefer a smaller chemical freeze-out temperature compared to strange hadrons~\cite{Bellwied:2013cta,Bellwied:2018tkc,Bluhm:2018aei}. It was also shown that the deconfinement crossover for the strange quarks may take place at a temperature larger than the critical temperature $T_{c}$~\cite{Ratti:2011au,HotQCD:2012fhj,Bazavov:2013dta}. The fluctuations of strangeness have also attracted the attention of many theoretical physicists ~\cite{Karsch:2010ck,Majumder:2006nq,Bollweg:2021vqf,Fu:2022gou,Ding:2020pao,Jin:2007wp,Koch:2005vg,Wang:2021wsj} who believe that it can reflect the strange sector of the QCD phase transition. In 2018, the STAR collaboration showed the first measurement for the moments of net-kaon multiplicity distributions in Au+Au collisions~\cite{STAR:2017tfy}, in which no nonmonotonic behavior was observed as a function of the colliding energy within experimental uncertainties. However, since net-kaon is not a conserved quantity, it is important to understand how its fluctuations and correlations go through the dynamic evolution stage of relativistic heavy-ion collisions~\cite{Zhou:2017jfk,Asakawa:2015ybt}.

In this study, we use a multiphase transport model (AMPT) to investigate the fluctuations of the net-kaon multiplicity distributions in Au+Au collisions at $\sqrt{s_{NN}} = 7.7$ GeV, and more importantly, to delve into the dynamical evolution of the cumulants and multiparticle correlations of the net-strangeness multiplicity distributions, following our previous works on net-charge~\cite{Huang:2021ihy} and net-proton fluctuations~\cite{Chen:2022wkj}. Our focus is on the extent to which the net-kaon fluctuations can be used as an approximation for the net-strangeness fluctuations in relativistic heavy-ion collisions. 

The paper is organized as follows. In Sec. II, we briefly introduce the AMPT model and show how we calculate the cumulants and correlation functions for the net-strangeness and net-kaon multiplicity distributions. In Sec. III, we compare our model results with the measurements from the STAR experiment, then present and discuss the results on the stage evolution of cumulants and multi-particle correlations of net-kaon and net-strangeness. A summary is given in Sec. IV.

\section{Model and calculation method}
\label{framework}

\subsection{The AMPT model}
\label{ampt}
AMPT~\cite{Lin:2004en} is extensively used to study relativistic heavy-ion collisions~\cite{Ma:2016fve,Ma:2011uma,Ma:2013pha,Bozek:2015swa,Bzdak:2014dia}. The AMPT model with a string melting mechanism is a hybrid model which includes four software packages to simulate four main stages in relativistic heavy-ion collisions. It uses a heavy ion jet interaction generator (HIJING) to simulate the initial condition, providing mainly the spatial and momentum distributions of minijet partons from the QCD hard process and soft string excitations~\cite{Wang:1991hta,Gyulassy:1994ew}. Many parent hadrons, including strange mesons and strange baryons, are temporarily produced by the fragmentation of minijet partons and excited strings~\cite{Sjostrand:2006za}. To mimic a partonic matter, these parent hadrons are melted into primordial quarks and antiquarks according to the flavor and spin structures of constitute quarks, using the string melting mechanism. The partonic interactions inside the formed quark plasma are simulated by Zhang's parton cascade (ZPC) model~\cite{Zhang:1997ej}, which currently only includes two-body elastic scattering using a perturbative QCD cross section (3 mb). When all partons stop interacting, the hadronization process is accomplished by a new quark coalescence model~\cite{He:2017tla}. A relativistic transport (ART) model~\cite{Li:1995pra} is used to simulate hadron rescatterings for the evolution of hadronic phase. In order to compare with experimental data, our simulations also include resonance decays, especially those of unstable strange hadrons (e.g., $\phi$, $\Lambda$, $\Sigma$, $\Xi$, and $\Omega$) which provide feed-down contributions to $\pi$ and $K$. It is worth mentioning that the AMPT model can give reasonable chemical freeze-out temperatures in agreement with the experimental measurement~\cite{Yu:2014epa,Xu:2017akx}.

In this work, we use a charge conservation version of the AMPT model with a new quark coalescence mechanism to collect a total of 70 million events to study cumulants and correlations of net-kaon and net-strangeness in Au+Au collisions at $\sqrt{s_{NN}} = 7.7$ GeV. Two key improvements in this version are introduced as follows:

i) This version of the AMPT model ensures the conservation of various conserved charges (including electric charge, baryon number, and strangeness) for all hadronic reaction channels during the evolution of hadronic phase. This version has been used to study the fluctuations of protons and baryons in Au+Au collisions at $\sqrt{s_{NN}} = 7.7$ GeV, which showed that the AMPT results on cumulants, cumulant ratios, correlation functions, and normalized correlation functions of proton, antiproton, and net-proton multiplicity distributions can basically describe the trend of the STAR data~\cite{Chen:2022wkj}. Furthermore, this version has given a good description of the STAR measured energy dependence of net-charge fluctuations~\cite{Huang:2021ihy}. We emphasize that the charge conservation version is necessary and important for the study of fluctuations of conserved charges, especially for strangeness, since the old version violates the conservation of conserved charges, especially for strangeness, due to the neutral kaon problem~\cite{Chen:2022wkj}.

ii) The treatment of hadronization process has been improved by introducing a new quark coalescence model ~\cite{He:2017tla}. The new quark coalescence model allows a quark to form either a meson or a baryon depending on the distance to its coalescence partner(s) (see Ref.~\cite{He:2017tla} for details please), rather than simply combining nearest quarks into hadrons in the old quark coalescence model~\cite{Lin:2004en}. It is important to use the new quark coalescence model to study the fluctuations of strangeness for the following reasons. It has been shown that the string-melting AMPT model with the new quark coalescence mechanism can provide a better description of the properties of bulk matter, especially for the production of strange hadrons, in relativistic heavy-ion collisions~\cite{He:2017tla,Lin:2021mdn}. On the other hand, the numbers of net-charge, net-baryon, and net-strangeness are conserved in the new quark coalescence model, instead of unnecessary separate conservation in the old coalescence model~\cite{He:2017tla,Lin:2021mdn}.

\subsection{Calculation method}
\label{sec:partB}
In statistics, various features of the probability distribution can be characterized by different moments, such as mean ($M$), variance ($\sigma ^{2}$), skewness ($S$), kurtosis ($\kappa$), as well as in terms of cumulants $C_{n}$. The various orders of cumulants $C_{n}$ of multiplicity distributions can be calculated as follows~\cite{Luo:2011rg,Luo:2010by,Luo:2011ts,Luo:2017faz}:
\begin{eqnarray}
C_{1}&=&\left \langle N \right \rangle, C_{2}=\left \langle \left (\delta N  \right )^{2} \right \rangle, C_{3}=\left \langle \left (\delta N  \right )^{3} \right \rangle,\notag \\ 
C_{4}&=&\left \langle \left (\delta N  \right )^{4} \right \rangle - 3\left \langle \left (\delta N  \right )^{2} \right \rangle^{2},\label{MDIV1}
\end{eqnarray}%
where $N$ is the number of particles on an event-by-event bias, $\delta N=N-\left \langle N \right \rangle$, and $\left \langle \cdots \right \rangle$ represents an event average. According to the definition of cumulants, various moments can be obtained as follows:
\begin{eqnarray}
M&=&C_{1}=\left \langle N \right \rangle, \sigma^{2}=C_{2}=\left \langle \left (\delta N  \right )^{2} \right \rangle,\notag \\ 
S&=&\frac{C_{3}}{\left ( C_{2} \right )^{\frac{3}{2}}}=\frac{\left \langle \left (\delta N  \right )^{3} \right \rangle}{\sigma ^{3}},\notag \\ \kappa&=&\frac{C_{4}}{(C_{2})^{2}}=\frac{\left \langle \left (\delta N  \right )^{4} \right \rangle}{\sigma ^{4}}-3.
\label{MDIV2}	
\end{eqnarray}%
In addition, the moment products can be expressed in terms of the ratios of cumulants, since these cumulant ratios are independent of volume and directly related to the susceptibilities of conserved charges~\cite{Ding:2015ona}:
\begin{eqnarray}
\frac{C_{2}}{C_{1}}=\frac{\sigma ^{2}}{M},\frac{C_{3}}{C_{2}}=S\sigma ,\frac{C_{4}}{C_{2}}=\kappa \sigma ^{2}.\label{MDIV3}
\end{eqnarray}

Let us consider a probability distribution function $P(N)$  to characterize the multiplicity distributions of particles in the final state, where $N$ is the number of only one species of particles, such as $K^{+}$ or $K^{-}$. By definition, $0\leq  P(N)\leq 1$. It is convenient to define a generating function~\cite{Bzdak:2016sxg,Bzdak:2019pkr}
\begin{eqnarray}
H_{\left ( z \right )}=\sum_{N}P\left ( N \right )z^{N},H_{\left ( 1 \right )}=1,
\label{MDIV6}	
\end{eqnarray}%
where the value of $H_{\left ( z \right )}$ at $z = 1$ is determined by the normalization condition $\sum_{N}P\left ( N \right )=1$. The factorial moment is conveniently calculated($i\geq 1$) as follow~\cite{Bzdak:2016sxg,Bzdak:2019pkr}:
\begin{eqnarray}
F_{i}=\left \langle \frac{N!}{\left ( N-i \right )!} \right \rangle=\frac{d^{i}}{dz^{i}}H\left ( z \right )\mid _{z=1}.
\label{MDIV7}	
\end{eqnarray}%
And the factorial cumulant (also known as the integrated correlation function) is given by  derivatives from the logarithm of $H_{\left ( z \right )}$~\cite{Bzdak:2016sxg,Bzdak:2019pkr},
\begin{eqnarray}
\kappa _{i}=\frac{d^{i}}{dz^{i}}\ln\left [ H\left ( z \right ) \right ]\mid _{z=1}.
\label{MDIV8}	
\end{eqnarray}%
The correlation functions, $\kappa _{n}$, can be expressed as follows in terms of the factorial moments:
\begin{eqnarray}
\kappa _{1}&=&F_{1},\notag \\
\kappa _{2}&=&F_{2}-F_{1}^{2},\notag \\
\kappa _{3}&=&F_{3}-3F_{1}F_{2}+2F_{1}^{3},\notag \\
\kappa _{4}&=&F_{4}-4F_{1}F_{3}-3F_{2}^{2}+12F_{1}^{2}F_{2}-6F_{1}^{4}.
\label{MDIV9}	
\end{eqnarray}%
Performing straightforward calculations, we can obtain the following cumulants based on the factorial moments:
\begin{eqnarray}
C_{2}&=&\kappa _{2}+\kappa _{1},\notag \\
C_{3}&=&\kappa _{3}+3\kappa _{2}+\kappa _{1},\notag \\      
C_{4}&=&\kappa _{4}+6\kappa _{3}+7\kappa _{2}+\kappa _{1}.
\label{MDIV10}
\end{eqnarray}%
Therefore, the cumulants are very useful since they are directly related to correlation functions. However, the cumulants also have a disadvantage because they mix different orders of correlation functions.

Another problem with the relationship of Eq.~(\ref{MDIV10}) is that it only works for analyzing one kind of particle. For example, to calculate the cumulants of net-kaon ($\Delta N_{K}=N^{K^{+}}-N^{K^{-}}$), two species of particles ($K^{+}$ and $K^{-}$) need to be considered, and the factorial moment can be expressed as $F_{i,k}=\left \langle \frac{N!}{\left ( N-i \right )!}  \frac{\bar{N}!}{\left ( \bar{N}-k \right )!} \right \rangle$, where $i$ is the number of $K^{+}$ and $k$ is that of $K^{-}$. It has been deduced that the relationship between the factorial moments and the correlation functions is satisfied as follows~\cite{Bzdak:2016sxg}:

\begin{eqnarray}
\kappa_{2}^{\left ( 2,0 \right )}&=&-F_{1,0}^{2}+F_{2,0},\notag \\
\kappa_{2}^{\left ( 1,1 \right )}&=&-F_{0,1}F_{1,0}+F_{1,1},\notag \\
\kappa_{3}^{\left ( 3,0 \right )}&=&2F_{1,0}^{3}-3F_{1,0}F_{2,0}+F_{3,0},\notag \\
\kappa_{3}^{\left ( 2,1 \right )}&=&2F_{0,1}F_{1,0}^{2}-2F_{1,0}F_{1,1}-F_{0,1}F_{2,0}+F_{2,1},\notag \\
\kappa_{4}^{\left ( 4,0 \right )}&=&-6F_{1,0}^{4}+12F_{1,0}^{2}F_{2,0}-3F_{2,0}^{2}-4F_{1,0}F_{3,0}+F_{4,0},\notag \\
\kappa_{4}^{\left ( 3,1 \right )}&=&-6F_{0,1}F_{1,0}^{3}+6F_{1,0}^{2}F_{1,1}+6F_{0,1}F_{1,0}F_{2,0}-3F_{1,1}F_{2,0}\notag \\
& &-3F_{1,0}F_{2,1}-F_{0,1}F_{3,0}+F_{3,1},\notag \\
\kappa_{4}^{\left ( 2,2 \right )}&=&\left ( -6F_{0,1}^{2}+2F_{0,2}\right )F_{1,0}^{2}+8F_{0,1}F_{1,0}F_{1,1}-2F_{1,1}^{2}\notag \\
& &-2F_{1,0}F_{1,2}+\left ( 2F_{0,1}^{2}-F_{0,2}\right )F_{2,0}-2F_{0,1}F_{2,1}+F_{2,2},
\label{MDIV11}
\end{eqnarray}%
where $\kappa_{n}^{\left ( i,k \right )}$ is an $n$-particle correlation function ($n=i+k$). The remaining correlation functions $\kappa_{n}^{\left ( i,k \right )}$ for $k> i$ can be easily obtained by a simple change of indexes $F_{i,k}\rightarrow F_{k,i}$. If both
$i$ and $k$ are not equal to 0, it means a mixed correlation function. For example, if $i$ and $k$ denote the number of $K^{+}$ (or $\bar{s}$ quarks) and $K^{-}$ (or $s$ quarks), $\kappa_{n}^{\left ( i,k \right )}$ represents the $n$-particle correlation between $i$ $K^{+}$ (or $\bar{s}$ quarks) and $k$ $K^{-}$ (or $s$ quarks). If 
$i$ or $k$ is equal to zero, it means a pure correlation function for one species of particles. For example, $\kappa_{n}^{\left ( i,0 \right )}$ and $\kappa_{n}^{\left ( 0,k \right )}$ represent the $i$-particle correlation for $K^{+}$ (or $\bar{s}$ quarks) and $k$-particle correlation for $K^{-}$ (or $s$ quarks), respectively. The relationship between the cumulants and the above correlation functions has been deduced as follows~\cite{Bzdak:2016sxg,Bzdak:2019pkr,Lin:2017xkd}:
\begin{eqnarray}:
C_{2}&=&\left \langle N \right \rangle+\left \langle \bar{N} \right \rangle+\kappa _{2}^{\left (2,0  \right )}+\kappa _{2}^{\left (0,2  \right )}-2\kappa _{2}^{\left (1,1  \right )},\notag \\
C_{3}&=&\left \langle N \right \rangle-\left \langle \bar{N} \right \rangle+3\kappa _{2}^{\left (2,0  \right )}-3\kappa _{2}^{\left (0,2  \right )}+\kappa _{3}^{\left (3,0  \right )}\notag \\
& &-\kappa _{3}^{\left (0,3  \right )}-3\kappa _{3}^{\left (2,1  \right )}+3\kappa _{3}^{\left (1,2  \right )},\notag \\
C_{4}&=&\left \langle N \right \rangle+\left \langle \bar{N} \right \rangle+7\kappa _{2}^{\left (2,0  \right )}+7\kappa _{2}^{\left (0,2  \right )}-2\kappa _{2}^{\left (1,1  \right )}\notag \\
& &+6\kappa _{3}^{\left (3,0  \right )}+6\kappa _{3}^{\left (0,3  \right )}-6\kappa _{3}^{\left (2,1  \right )}-6\kappa _{3}^{\left (1,2  \right )}+\kappa _{4}^{\left ( 4,0 \right )}\notag \\
& &+\kappa _{4}^{\left ( 0,4 \right )}-4\kappa _{4}^{\left ( 3,1 \right )}-4\kappa _{4}^{\left ( 1,3 \right )}+6\kappa _{4}^{\left ( 2,2 \right )}. 
\label{MDIV12}
\end{eqnarray}%

In our work, we apply the same kinematic cuts~\cite{STAR:2017tfy} as in the STAR experiment to calculate the various cumulants of the net-kaon multiplicity distributions. For example, we select $K^{+}$ and $K^{-}$ within a transverse momentum range of $0.2<p_{T}<1.6$ GeV/$c$ and a midrapidity window of $\left | y \right |< 0.5$. We adopt the charged particle multiplicity distributions to divide the centrality bins. In order to avoid self-correlation, the charged particle multiplicity other than $K^{+}$ and $K^{-}$ within pseudorapidity $\left |\eta   \right |< 1$ is used. We apply the bootstrap method to estimate the statistical error~\cite{Luo:2017faz}. To suppress the volume fluctuations caused by the finite central bin width and initial volume (geometric) fluctuations of the colliding nuclei~\cite{Luo:2011ts,Luo:2013bmi,He:2018mri,Huang:2021ihy,Chen:2010kqv}, we apply the central bin width correction (CBWC) in the calculations of cumulants and correlation functions~\cite{Chen:2022wkj,Huang:2021ihy,STAR:2021iop,Luo:2017faz}. Both analyses of STAR data and AMPT data showed that CBWE contributes significantly to higher-order cumulants, especially for more central collisions~\cite{Chen:2022wkj}.

In our analysis of stage evolution, we also focus on the net-strangeness fluctuations. For calculating net-strangeness, $N^{\rm nets}=N^{\bar{s}}-N^{s}$, $N^{\bar{s}}$ and $N^{s}$ represent the number of $\bar{s}$ quarks and $s$ quarks in one event. Note that the strangeness quantum number carried by $\bar{s}$ quark and $s$ quark is positive one and negative one, respectively. In the final state of heavy-ion collisions, different strange hadrons have different numbers of (anti) strange quarks. For example, $\Sigma$, $\Xi$, and $\Omega$ consist of 1, 2, and 3 strange quarks, respectively. The number of strange quarks in an event should be expressed as $N^{s}=\sum_{i}n_{i}^{s}p_{i}^{s}$, where $n_{i}^{s}$ represents the number of constitute $s$ quarks and $p_{i}$ represents the number of the strange hadron $i$. And vice versa, the number of constitute $\bar{s}$ quarks is $N^{\bar{s}}=\sum_{i}n_{i}^{\bar{s}}p_{i}^{\bar{s}}$. 

\section{Results and Discussions}
\label{framework}

\subsection{Centrality dependence}
\label{sec:partC}

\begin{figure}[htb]
\centering
\includegraphics
[width=9cm]{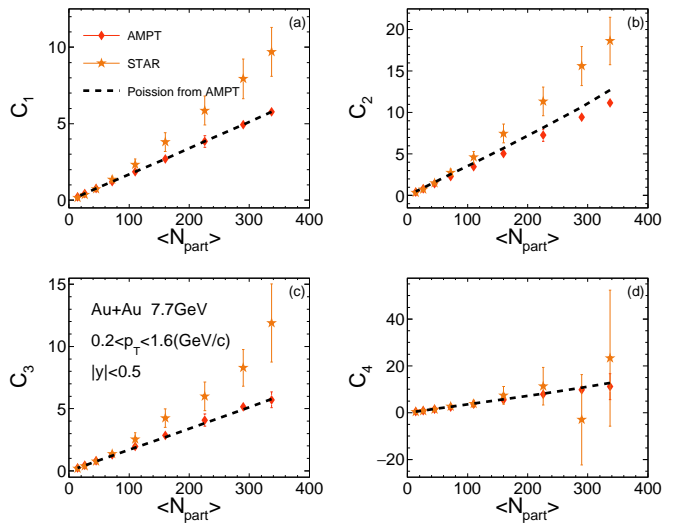}
\caption{(Color online) The AMPT results on cumulants $C_{n}$ of net-kaon ($\Delta N_{K}$) multiplicity distributions as a function of $\left \langle N_{part} \right \rangle$ in Au+Au collisions at $\sqrt{s_{NN}} = 7.7$ GeV, in comparisons with the STAR measurements~\cite{STAR:2017tfy}. The dotted line represents the Poisson baseline calculated using the mean particle multiplicity from the AMPT model.}
\label{FIG.1.}
\end{figure}
\begin{figure*}[htb]
\centering
\includegraphics
[width=17.5cm]{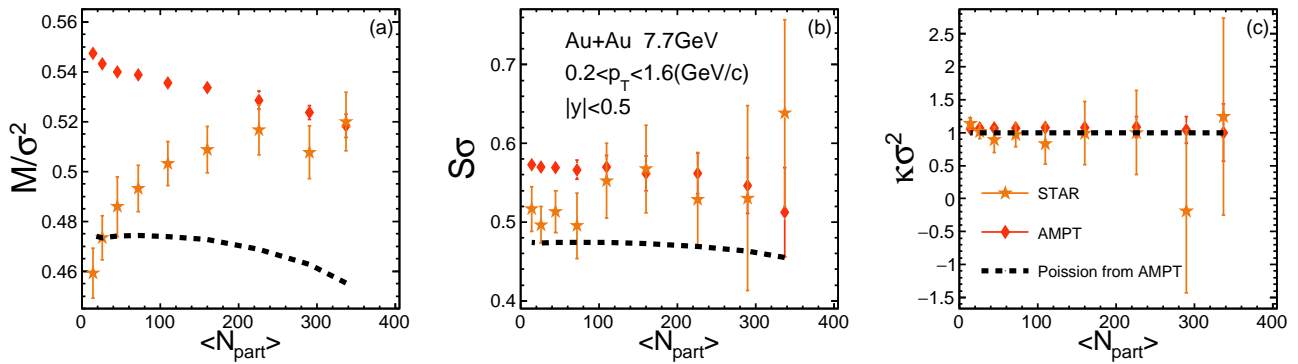}
\caption{(Color online) The AMPT results on cumulant ratios of net-kaon ($\Delta N_{K}$) multiplicity distributions as a function of $\left \langle N_{part} \right \rangle$ in Au+Au collisions at $\sqrt{s_{NN}} = 7.7$ GeV, in comparisons with the STAR measurements~\cite{STAR:2017tfy}. The dotted line represents the Poisson baseline calculated using the mean particle multiplicity from the AMPT model.}
\label{FIG.2.}
\end{figure*}

Figure~\ref{FIG.1.} shows the $\left \langle N_{part} \right \rangle$ (the average number of participant nucleons) dependences of cumulants $C_{n}$ ($n=1$, 2, 3, and 4) of net-kaon ($\Delta N_{K}$) multiplicity distributions in Au+Au collisions at $\sqrt{s_{NN}} = 7.7$ GeV. The cumulants $C_{n}$ of net-kaon increase monotonically from peripheral to central collisions. We found that the AMPT results are all slightly smaller than the experimental data. Assuming that the multiplicity distributions of $K^{+}$ and $K^{-}$ are independent Poisson distributions, the corresponding net-kaon Poisson expectation can be obtained by $C_{n}^{\rm Poisson}=C_{n}^{K^{+}}-\left ( -1 \right )^{n}C_{n}^{K^{-}}$. Within statistical uncertainties, all values of the cumulant $C_{n}$ seem to be consistent with the Poisson baseline calculated using the mean particle multiplicity from the AMPT model. However, if you look closely at Fig.~\ref{FIG.1.}(b), the $C_{2}$ for the AMPT model is slightly lower than the Poisson baseline based on its mean multiplicity, suggesting a correlation between $K^{+}$ and $K^{-}$, which will be discussed later.

\begin{figure*}[htb]
\centering
\includegraphics
[width=17.5cm]{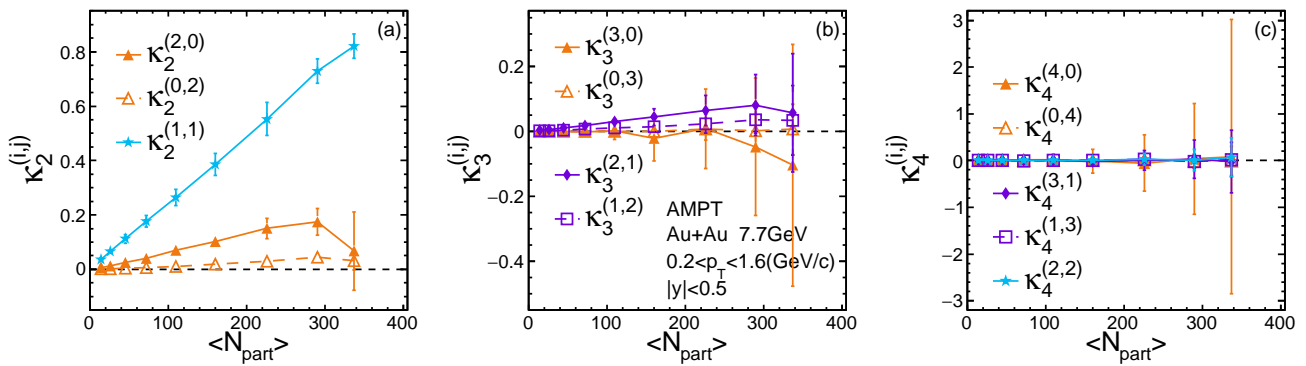}
\caption{(Color online) The AMPT results on correlation functions of $K^{+}$ and $K^{-}$ as a function of $\left \langle N_{part} \right \rangle$ in Au+Au collisions at $\sqrt{s_{NN}} = 7.7$ GeV.}
\label{FIG.3.}
\end{figure*}
 
Figure~\ref{FIG.2.} shows the $\left \langle N_{part} \right \rangle$ dependences of cumulant ratios [$C_{1}/C_{2}\left ( M/\sigma ^{2} \right )$, $C_{3}/C_{2}\left ( S\sigma \right )$, and $C_{4}/C_{2}\left ( \kappa \sigma^{2} \right )$] of net-kaon ($\Delta N_{K}$) multiplicity distributions in Au+Au collisions at $\sqrt{s_{NN}} = 7.7$ GeV. In Fig.~\ref{FIG.2.}(a), the AMPT result on $M/\sigma ^{2}$ can describe the experimental data for central collisions, however, the overall trend is opposite to the experimental trend due to its overestimation for peripheral collisions. We have checked that this is caused by the new quarks coalescence mechanism by comparing the two versions of AMPT models with old and new quark coalescence mechanisms, which will be discussed later. The $M/\sigma ^{2}$ for the AMPT model is also larger than the Poisson baseline based on its mean multiplicity. In Fig.~\ref{FIG.2.}(b), the AMPT result is consistent with the experimental measurement of $S\sigma$ of net-kaon for nonperipheral collisions within the uncertainties, and larger than the Poisson baseline based on its mean multiplicity. The AMPT result for $\kappa \sigma^{2}$ of net-kaon is consistent with both the experimental measurement and the Poisson baseline based on its mean multiplicity within the large errors, as shown in Fig.~\ref{FIG.2.}(c). These differences between the net-kaon moment products and the Poisson expectations indicate that the net-kaon is not just a simple random or independent case, and there must be some correlations between the kaon particles. 

To see the multiparticle correlations among kaons, Fig.~\ref{FIG.3.} shows the $\left \langle N_{part} \right \rangle$ dependences of correlation functions of $K^{+}$ and $K^{-}$ multiplicity distributions in Au+Au collisions at $\sqrt{s_{NN}} = 7.7$ GeV, which are obtained according to Eq.~(\ref{MDIV11}). In Fig.~\ref{FIG.3.}(a), we observe that all three kinds of two-particle correlations for $K^{+}$ and $K^{-}$ are positive. The opposite-charge two-particle correlation [$\kappa _{2}^{\left (1,1  \right )}$] between $K^{+}$ and $K^{-}$ is significantly greater than the same-charge two-particle correlations [$\kappa _{2}^{\left (2,0  \right )}$ and $\kappa _{2}^{\left (0,2  \right )}$]. It indicates that the pair production of strangeness is likely to be a dominant source, which will be discussed in detail in the next subsection. We have found that the opposite trend of the $\left \langle N_{part} \right \rangle$ dependence of $M/\sigma ^{2}$ may be due to the slope (growth rate) of $\kappa _{2}^{\left (1,1  \right )}$ in peripheral collisions being too large. A possible solution is that ones can introduce a centrality dependence of $\kappa _{2}^{\left (1,1  \right )}$ in the new coalescence scheme. To achieve it, we refer to Ref.~\cite{Shao:2020sqr} where they introduced the parameters $\gamma _{\Lambda ,\Xi ,\Omega }$ in the coalescence scheme to control the yields of the interested particles. Similarly, we may introduce the parameters $\gamma _{K^{+} }$ and $\gamma _{K^{-} }$ to make $K^{+}$ and $K^{-}$ less correlated with a desired centrality dependence of $\kappa _{2}^{\left (1,1  \right )}$. We would like to leave it as our future study. On the other hand, the three- and four-particle correlation functions for $K^{+}$ and $K^{-}$ are consistent with zero within the current statistical uncertainties, as shown in Figs.~\ref{FIG.3.}(b) and (c), respectively.

\newcommand{\tabincell}[2]{\begin{tabular}{@{}#1@{}}#2\end{tabular}}
\begin{table}[htbp]
\centering
\caption{\label{comparison} Five evolution stages in the AMPT model with a string melting mechanism.}
\renewcommand\arraystretch{1.5} 
\scalebox{.75}{
\begin{tabular}{lc}\\
\hline  \hline
Stage& Description\\
\hline
(a) Initial state & \tabincell{c}{The initial state of partonic matter consisting\\ of quarks and antiquarks created \\by $A+A$ collisions}\\
\hline
(b) After parton cascade & \tabincell{c}{The final state of partonic matter consisting\\ of quarks and antiquarks that have\\ undergone parton cascade} \\
\hline
(c) After hadronization & \tabincell{c}{The initial state of hadronic matter transformed\\ from the freeze-out partonic matter through\\ the hadronization of coalescence}\\
\hline
(d) After hadronic rescatterings & \tabincell{c}{The freeze-out hadronic matter which has \\ undergone hadronic rescatterings} \\
\hline
(e) Final state & \tabincell{c}{The final state of hadronic matter with\\ considering weak decays} \\
\hline  \hline
\end{tabular}
}
\label{TABLE}
\end{table}

\subsection{Stage evolution}
\label{sec:partC}

Relativistic heavy-ion collisions are actually a complex dynamical evolution, including several important evolution stages. In order to better understand the dynamics of fluctuation observables, it is necessary to study the cumulants and correlation functions for each evolution stage. With the help of the AMPT model, we focus on the dynamic development of the cumulants and correlation functions for the five evolution stages in Au+Au collisions at $\sqrt{s_{NN}} = 7.7$ GeV using the AMPT model. The five evolution stages (a)--(e) are described in Table~\ref{TABLE}.

\begin{figure*}[htbp]
\centering
\includegraphics
[width=17.5cm]{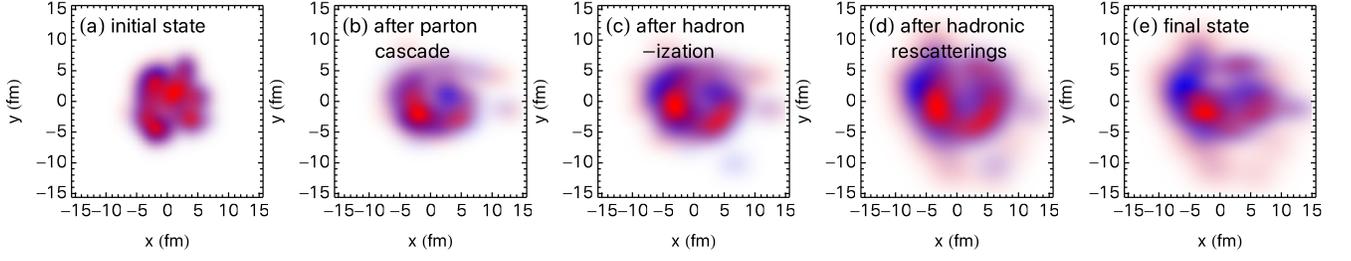}
\caption{The AMPT results on the density distributions of $s$ (constitute) quarks and $\bar{s}$ (constitute) quarks in midrapidity at different evolution stages for a selected central Au+Au collision event at $\sqrt{s_{NN}} = 7.7$ GeV, where red and blue colors represent $s$ and $\bar{s}$, respectively.}
\label{FIG.4.}
\end{figure*}

Since the conserved quantity is not net-kaon but net-strangeness, we will focus on the dynamic development of the cumulants and correlation functions for strangeness next. To calculate the cumulants of the net-strangeness distribution for the last three hadronic stages, we count the numbers of constitute $s$ and $\bar{s}$ quarks inside the strange hadrons and their antiparticles, including $K^{-}$, $\bar{K^{0}}$, $\Lambda$, $\Sigma$, $\Xi$, $\Omega$, etc. This is similar to the fourth case in Ref.~\cite{Zhou:2017jfk}. The number of strangeness arises and develops during the evolution stages, which also fluctuates event by event. Figure~\ref{FIG.4.} illustrates the density distributions of $s$ and $\bar{s}$ (constitute) quarks in midrapidity at different evolution stages for a selected AMPT event of central Au+Au collision at $\sqrt{s_{NN}} = 7.7$ GeV. Our task in this work is to study the stage evolution pattern of strangeness fluctuations. In addition, we also try to understand to what extent the net-kaon fluctuations can be regarded as net-strangeness fluctuations in relativistic heavy-ion collisions.

\begin{figure*}[htbp]
\centering
\includegraphics
[width=14.5cm]{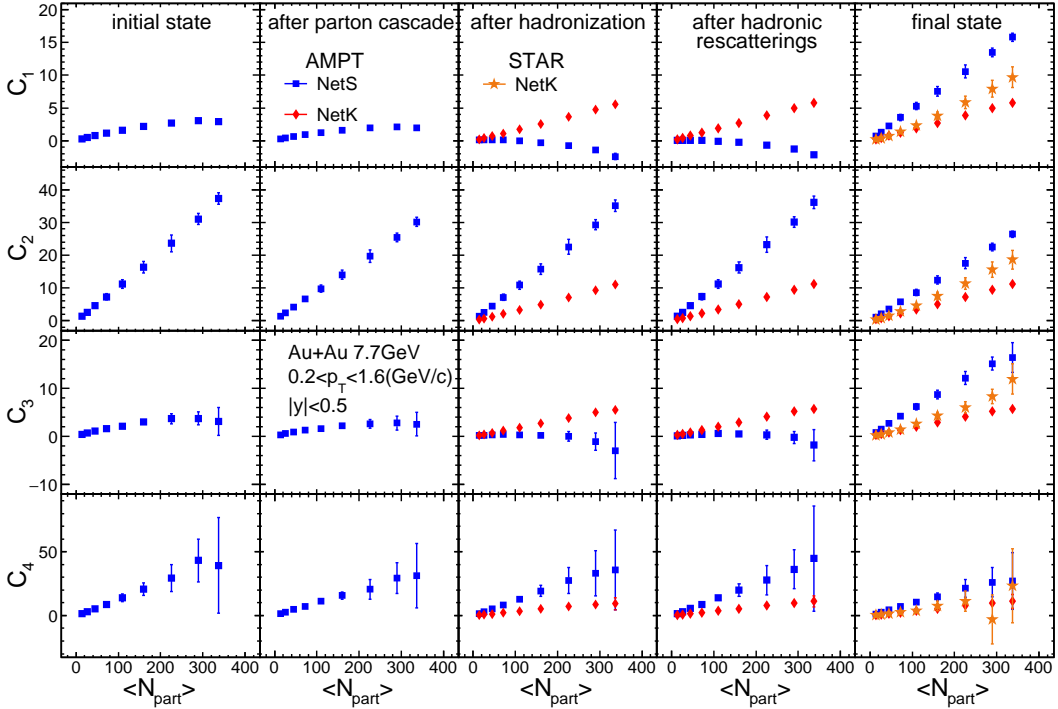}
\caption{The AMPT results on the cumulants $C_{n}$ of net-strangeness or net-kaon distributions as a function of $\left \langle N_{part} \right \rangle$ at different evolution stages in Au+Au collisions at $\sqrt{s_{NN}} = 7.7$ GeV, in comparisons with the STAR measurements~\cite{STAR:2017tfy}.}
\label{FIG.5.}
\end{figure*}

 Figure~\ref{FIG.5.} shows the $\left \langle N_{part} \right \rangle$ dependences of cumulants $C_n$ ($n=1$,2,3, and 4) of net-strangeness multiplicity distributions at five different evolution stages (from left to right) in Au+Au collisions at $\sqrt{s_{NN}} = 7.7$ GeV. First, let us focus on $C_1$ for the moment, because For $C_n$ ($n=2$,3, and 4), they include complex contributions from multiple-particle correlations as illustrated by Eq.~(\ref{MDIV12}), which will be discussed with Fig.~\ref{FIG.7.} in details later. Next, let us only focus on the first four columns and ignore the last column, because the first four evolution stages strictly obey the law of conservation of strangeness, while the last evolution stage contains weak decays that break the strangeness conservation law. From ``Initial state'' to ``After hadronic rescatterings'', it is observed that the first-order cumulant $C_{1}$ for net-strangeness are close to zero. It is understandable since the initial participant nucleons have no strange hadrons and the total net-strangeness should be zero. However, since we only focus on the mid-rapidity window with a transverse momentum cut, this leads to some deviations from zero. On the other hand, the first-order cumulant $C_{1}$ for net-kaon is clearly positive because the strange hadrons include not only kaons but also other strange hadrons like many strange baryons. For the final state in the last column which experiences weak decays, it is found that the mean($C_{1}$) value of the net-strangeness becomes significantly enhanced with respect to the result of after hadronic rescatterings. It can be understood by the following example. For the decay of $\Lambda \rightarrow \pi ^{-}+p$, $\Lambda$ carries $s$ quarks, but no $s$ quarks inside the decay daughters after weak decays. In addition, the yield of strange baryons is greater than that of antistrange baryons at low energies. Therefore, the reduction of the $s$ quark should be more than the $\bar{s}$ quark, which leads to the enhancement of the number of net-strangeness after weak decays. Fortunately, we find that the net-kaon $C_{n}$ keeps unchanged in the last three stages, which indicates that both hadronic rescatterings and resonance decays have little effect on the net-kaon fluctuations. 

\begin{figure*}[htbp]
\centering
\includegraphics
[width=14.5cm]{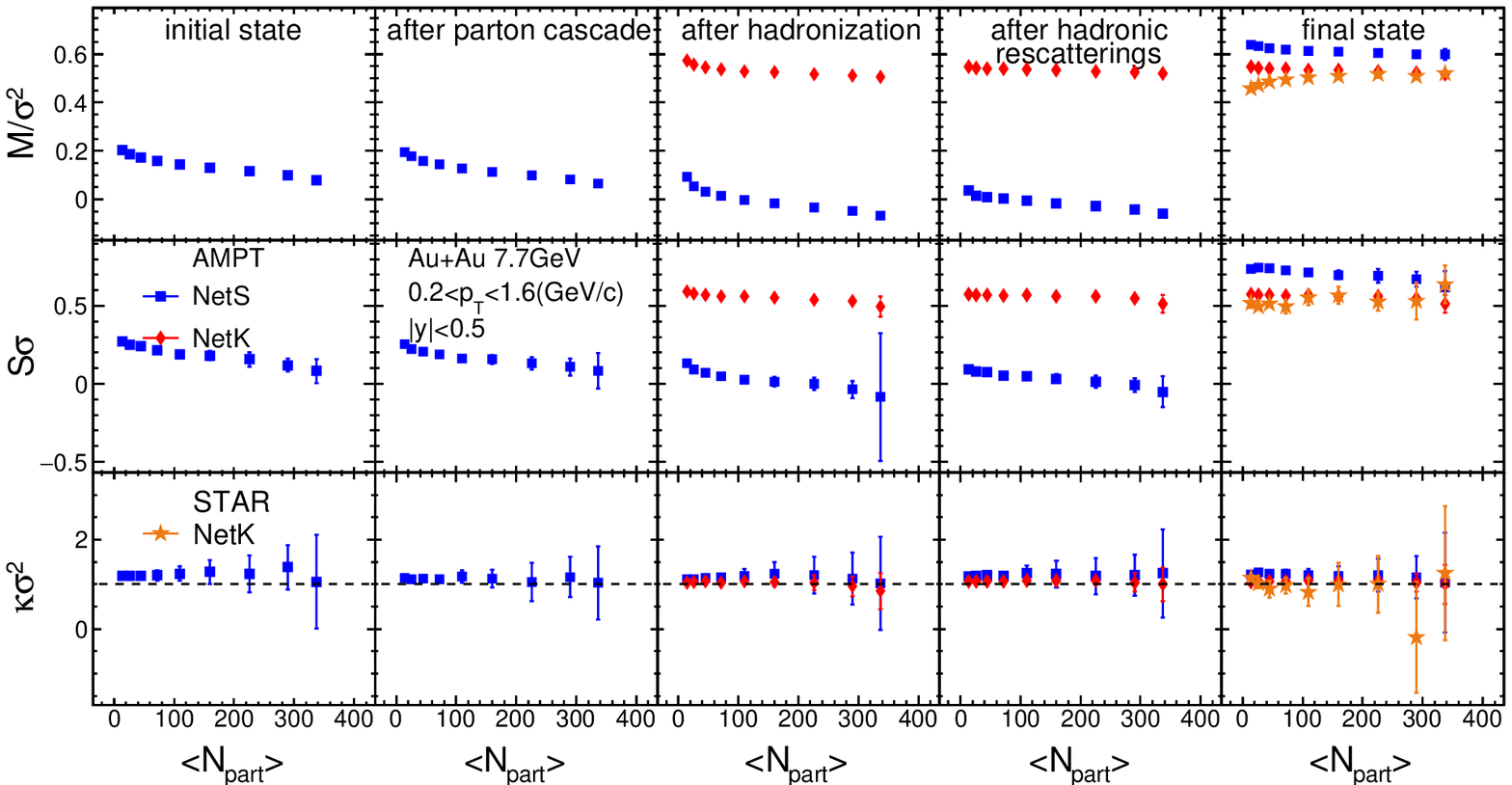}
\caption{The AMPT results on cumulant ratios of net-strangeness or net-kaon distributions as a function of $\left \langle N_{part} \right \rangle$ at different evolution stages in Au+Au collisions at $\sqrt{s_{NN}} = 7.7$ GeV, in comparisons with the STAR measurements~\cite{STAR:2017tfy}.}
\label{FIG.6.}
\end{figure*}

Figure~\ref{FIG.6.} shows the $\left \langle N_{part} \right \rangle$ dependence of cumulant ratios of net-strangeness multiplicity distributions at five different evolution stages in Au+Au collisions at $\sqrt{s_{NN}} = 7.7$ GeV. It can be found that the magnitudes of $M/\sigma ^{2}$ and $S\sigma$ for net-strangeness gradually decrease from ``Initial state" to ``after hadronic rescatterings" with the stage evolution of heavy-ion collisions. As $\left \langle N_{part} \right \rangle$ increases, $M/\sigma ^{2}$ and $S\sigma$ decrease or even become negative. For the final state in the last column, we observe that the values of $M/\sigma ^{2}$ and $S\sigma$ for net-strangeness both increase suddenly due to weak decays. The values of $\kappa \sigma^{2}$ for net-strangeness are consistent with unity within errors. The net-kaon results show similar trends to net-strangeness, but with slightly different magnitudes. We observe that the cumulant ratios of net-kaon multiplicity distributions also keep unchanged for the last three stages, since we have seen that the net-kaon cumulants hardly suffer from hadronic rescatterings and resonance decays. We conclude that the net-kaon fluctuations can represent the net-strangeness fluctuations, even with slightly different magnitudes. Therefore, it is a good way to use net-kaon fluctuations to search for possible QCD critical fluctuations of net strangeness predicted by lattice QCD~\cite{Cheng:2008zh} and effective QCD models~\cite{Fu:2009wy,Chahal:2022syd}. 

\begin{figure*}[htbp]
\centering
\includegraphics
[width=14.5cm]{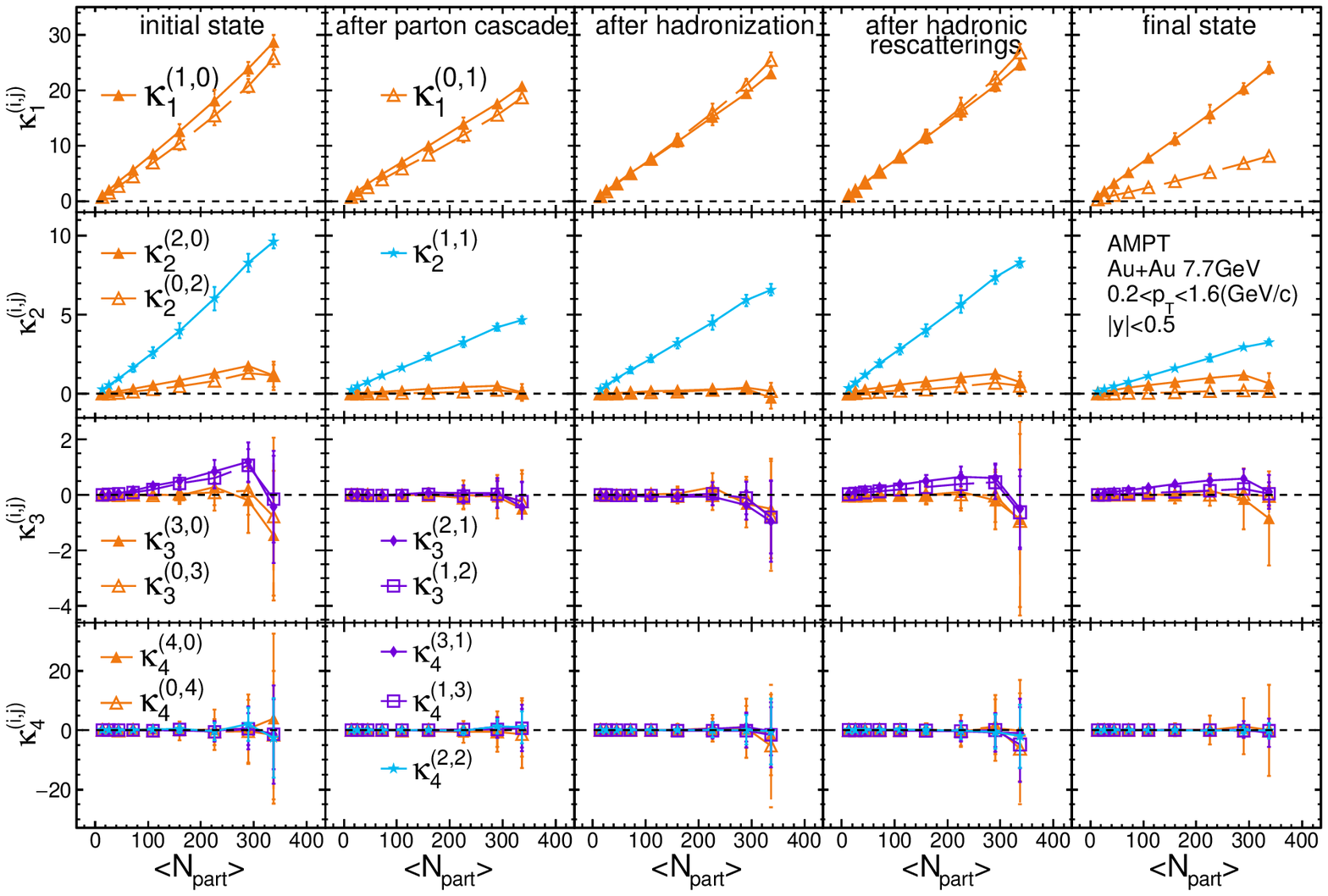}
\caption{The AMPT results on $n$-particle correlation functions $\kappa _{n}^{\left ( i,j \right )}$ of $s$ (constitute) quarks and $\bar{s}$ (constitute) quarks as a function of $\left \langle N_{part} \right \rangle$ at different evolution stages in Au+Au collisions at $\sqrt{s_{NN}} = 7.7$ GeV. Note that the plots in the top row show the stage evolution of the mean particle number in fact.}
\label{FIG.7.}
\end{figure*}

 To further understand the physics mechanisms for the stage evolution of the cumulants of net-strangeness, Fig.~\ref{FIG.7.} shows the $\left \langle N_{part} \right \rangle$ dependences of $n$-particle correlation functions $\kappa _{n}^{\left ( i,j \right )}$ of $j$ $s$ (constitute) quarks and $i$ $\bar{s}$ (constitute) quarks ($n$=$i+j$) at five different evolution stages in Au+Au collisions at $\sqrt{s_{NN}} = 7.7$ GeV. Note that the plots in the top row actually show the stage evolution of the numbers of $\bar{s}$ and $s$ quarks, since $\kappa _{1}^{\left ( 1,0 \right )}$=$\left \langle N^{\bar{s}}\right \rangle$ and $\kappa _{1}^{\left ( 0,1 \right )}$=$\left \langle N^{s} \right \rangle$. We observe that the three- and four-particle correlation functions are consistent with zero throughout the evolution of heavy-ion collisions within our limited statistics.\footnote{There seem to be very small signals of $\kappa _{2}^{\left ( 2,1 \right )}$ and $\kappa _{2}^{\left ( 1,2 \right )}$, which may be caused by the production of two pairs of $s\bar{s}$ in the string fragmentation.} Next, we will focus only on the one-particle yields and two-particle correlation functions. For the initial state, we observe that the numbers of $\bar{s}$ and $s$ quarks are very close. All of the three two-particle correlation functions are positive, however, the two-particle correlation function between the $\bar{s}$ quark and $s$ quark [$\kappa _{2}^{\left ( 1,1 \right )}$] is dominant. It indicates that the $s$ and $\bar{s}$ quarks are produced in pairs due to the conservation law of strangeness, which is visualized by Fig.~\ref{FIG.4.}(a). The finite signals of same-charge two-particle correlations [$\kappa _{2}^{\left ( 2,0 \right )}$ and $\kappa _{2}^{\left ( 0,2 \right )}$] can be considered as a result of the melting of parent multistrange baryons under the string melting mechanism in the AMPT model. Note that the signs of two-particle correlations for strangeness differ from that for baryons~\cite{Chen:2022wkj}, because strangeness is newly created, however, baryon number arises from baryon stopping. After parton cascade, we find that the numbers of $\bar{s}$ and $s$ quarks decrease, and two-particle correlations are weakened. This is the result of the strangeness diffusion during the evolution of partonic phase, which is visualized by Fig.~\ref{FIG.4.}(b). It is similar to what we have found in our previous study on net-baryon fluctuations, in which the diffusion of baryon number also weakens the two-baryon correlation during the expansion of the fireball~\cite{Chen:2022wkj}. Since the hadronization converts quarks into hadrons and the kinematics of formed strange hadrons is different from that of constitute strange quarks, it leads to some enhancement of strangeness yields and two-particle correlation functions within the acceptance window, which is visualized by Fig.~\ref{FIG.4.}(c). After hadronic rescatterings, the strangeness yields and two-particle correlation functions also look increased, which is visualized by Fig.~\ref{FIG.4.}(d). This can be understood since some reactions can not only create two strange mesons with $s$ and  $\bar{s}$ quarks, but also a strange baryon carrying two $s$ quarks. For example, i) in the reaction channel $\pi^{0} + \pi^{0} \leftrightarrow K^{-} + K^{+}$, because $K^{-}$ and $K^{+}$ carry one $s$ quark and one $\bar{s}$ quark, respectively, their creation and annihilation will inevitably change $\kappa _{2}^{\left ( 1,1 \right )}$;  ii) in the reaction channel $\Lambda ^{0}+K^{-} \leftrightarrow \Xi ^{-}+\pi^{0}$, because $\Xi ^{-}$ carries two $s$ quarks, and the creation and annihilation of $\Xi ^{-}$ will inevitably modify $\kappa _{2}^{\left ( 0,2 \right )}$. For the final state, the number of $s$ quarks decreases because the decays of strange baryons destroy the existence of strangeness, which is visualized by Fig.~\ref{FIG.4.}(e). At the same time, the two-particle correlation functions, especially for $\kappa _{2}^{\left ( 1,1 \right )}$, are reduced. Therefore, the stage evolution of these different multiparticle correlations together contributes to the stage evolution of cumulants of net-strangeness shown in Fig.~\ref{FIG.5.} by Eq.~(\ref{MDIV12}). By comparing Fig.~\ref{FIG.3.} and ~\ref{FIG.7.}, we find that the centrality dependences of the correlation functions for kaons follow similar trends to those of strangeness in the final state, which supports that the fluctuations of net-kaon can be a good proxy for the fluctuations of net-strangeness.
 
\section{Summary}
\label{framework}

In summary, the centrality dependences of cumulants and correlation functions of net-kaon multiplicity distributions have been studied in Au+Au collisions at $\sqrt{s_{NN}} = 7.7$ GeV using a new version of the multiphase transport model with a new coalescence mechanism and all charge conservation laws. Our results of cumulants and cumulant ratios net-kaon multiplicity distributions qualitatively describe the STAR data in trends. By studying the stage evolution of the cumulants, cumulant ratios, and correlation functions, we reveal several key effects on the fluctuations and correlations of strangeness during the dynamical evolution of relativistic heavy-ion collisions. Our results indicate that the $s$ and $\bar{s}$ quarks are produced in pairs in the initial state. The produced $s$ and $\bar{s}$ quarks are diffused due to parton cascade. The correlation between $s$ and $\bar{s}$ quarks in midrapidity is enhanced by hadronization and hadronic rescatterings, however, it is weakened due to weak decays of strange baryons. Our results give a relatively complete description of the dynamics of strangeness in relativistic heavy-ion collisions. Without considering the QCD critical fluctuations, we find that the net-kaon fluctuations can largely represent the net-strangeness fluctuations because net-kaon follows similar trends to net-strangeness, independent of the evolution of the hadronic phase. However, since the physics of the QCD critical fluctuations is not included in the current AMPT model, our results are expected to only provide a helpful baseline for searching for the possible critical behaviors at the CEP in relativistic heavy-ion collisions. 

Our study also indicates that further improvements of the AMPT model are needed for studying the fluctuations of strangeness in heavy-ion collisions at beam energy scan energies, especially for the initial condition and the hadronization scheme. However, we have checked that the finite nuclear thickness effect~\cite{Wang:2021owa} has little influence on the cumulant ratios and correlation functions. Therefore, we hope that further improvements to the hadronization mechanism will help to describe the measured fluctuations of strangeness, which will be a direction for our future research.

\section*{ACKNOWLEDGMENTS}
We thank Profs. Xiao-Feng Luo，Weijie Fu and Mei Huang for helpful discussions, Prof. Zi-Wei Lin for providing the AMPT code with the new quark coalescence, and Dr. Chen Zhong for maintaining the high-quality performance of the Fudan supercomputing platform for nuclear physics. This work is supported by the National Natural Science Foundation of China under Grants No.12147101, 11890714, 11835002, 11961131011, 11421505, the National Key Research and Development Program of China under Grant No. 2022YFA1604900, the Strategic Priority Research Program of Chinese Academy of Sciences under Grant No. XDB34030000, and the Guangdong Major Project of Basic and Applied Basic Research under Grant No. 2020B0301030008.

\bibliography{myref}


\end{CJK*}

\end{document}